\documentclass{article}
\usepackage{spconf,amsmath,amsfonts,graphicx,url,stmaryrd}
\usepackage[justification=centering]{caption}
\graphicspath{{figures/}}

\title{Phase reconstruction of spectrograms with linear unwrapping: application to audio signal restoration}

\name{Paul Magron \qquad Roland Badeau \qquad Bertrand David}

\address{Institut Mines-T\'{e}l\'{e}com, T\'{e}l\'{e}com ParisTech, CNRS LTCI, Paris, France \\ \texttt{<firstname>.<lastname>@telecom-paristech.fr}}

\begin{document}

\maketitle

\begin{abstract}
This paper introduces a novel technique for reconstructing the phase of modified spectrograms of audio signals. From the analysis of mixtures of sinusoids we obtain relationships between phases of successive time frames in the Time-Frequency (TF) domain. To obtain similar relationships over frequencies, in particular within onset frames, we study an impulse model. Instantaneous frequencies and attack times are estimated locally to encompass the class of non-stationary signals such as vibratos. These techniques ensure both the vertical coherence of partials (over frequencies) and the horizontal coherence (over time). The method is tested on a variety of data and demonstrates better performance than traditional consistency-based approaches. We also introduce an audio restoration framework and observe that our technique outperforms traditional methods.
\end{abstract}


\section{Introduction}

A variety of music signal processing techniques act in the TF domain, exploiting the particular structure of music signals. For instance, the family of techniques based on Nonnegative Matrix Factorization (NMF) is often applied to spectrogram-like representations, and has proved to provide a successful and promising framework for source separation~\cite{Smaragdis2003}. Magnitude-recovery techniques are also useful for restoring missing data in corrupted signals~\cite{Fitzgerald2012}.

However, when it comes to resynthesizing time signals, the phase recovery of the corresponding Short-Time Fourier Transform (STFT) is necessary. In the source separation framework, a common practice consists in applying Wiener-like filtering (soft masking of the complex-valued STFT of the original mixture). When there is no prior on the phase of a component (\emph{e.g.} in the context of audio restoration), a consistency-based approach is often used for phase recovery~\cite{Griffin1984}. That is, a complex-valued matrix is iteratively computed to be close to the STFT of a time signal. A recent benchmark has been conducted to assess the potential of source separation methods with phase recovery in NMF~\cite{Magron2015}. It points out that consistency-based approaches provide poor results in terms of audio quality. Besides, Wiener filtering fails to provide good results when sources overlap in the TF domain. Thus, phase recovery of modified audio spectrograms is still an open issue. The High Resolution NMF (HRNMF) model~\cite{Badeau2014} has shown to be a promising approach, since it models a TF mixture as a sum of autoregressive (AR) components in the TF domain, thus dealing explicitly with a phase model.

Another approach to reconstruct the phase of a spectrogram is to use a phase model based on the observation of fundamental signals that are mixtures of sinusoids. Contrary to consistency-based approaches using the redundancy of the STFT, this model exploits the natural relationship between adjacent TF bins due to the model. This approach is used in the phase vocoder algorithm~\cite{Laroche1999}, although it is mainly dedicated to time stretching and pitch modification of signals, and it requires the phase of the original STFT.
%More recently,~\cite{Bronson2014} proposed a complex NMF framework with phase constraints based on sinusoidal modeling. Although promising, this approach is limited to harmonic and stationary signals, and requires prior knowledge on fundamental frequencies and numbers of partials.
More recently,~\cite{Bronson2014} proposed a complex NMF framework with phase constraints based on sinusoidal modeling, and~\cite{Krawczyk2014} used a similar technique for recovering the phase of speech signals in noisy mixtures. Although promising, these approaches are limited to harmonic and stationary signals. Besides, the phase constrained complex NMF model~\cite{Bronson2014} requires prior knowledge on fundamental frequencies and numbers of partials. In the speech enhancement framework introduced in~\cite{Krawczyk2014}, the fundamental frequency is estimated, however the estimation error is propagated and amplified through partials and time frames.

In this paper, we propose a generalization of this approach that consists in estimating the phase of mixtures of sinusoids from its explicit calculation. We then obtain an algorithm which unwraps the phases \emph{horizontally} (over time frames) to ensure the temporal coherence of the signal, and \emph{vertically} (over frequency channels) to enforce spectral coherence between partials, which is observed in musical acoustics for several instruments~\cite{Galembo2001}. Our technique is suitable for a variety of pitched music signals, such as piano or guitar sounds, but percussive signals are outside the scope of this research. A dynamic estimation (at each time frame) of instantaneous frequencies extends the validity of this technique to non-stationary signals such as cellos and speech. This technique is tested on a variety of signals and integrated in an audio restoration framework.

The paper is organized as follows. Section \ref{sec:horizontal} presents the horizontal phase unwrapping model. Section \ref{sec:onset} is dedicated to phase reconstruction on onset frames. Section \ref{sec:exp} presents a performance evaluation of this technique through various experiments. Section \ref{sec:inpainting} introduces an audio restoration framework using this phase recovery method. Finally, section \ref{sec:conclu} draws some concluding remarks.

\section{Horizontal phase reconstruction}

\label{sec:horizontal}

\subsection{Sinusoidal modeling}
Let us consider a sinusoid of normalized frequency $f_0 \in [-\frac{1}{2} ; \frac{1}{2}] $, initial phase $\phi_0 \in [-\pi ; \pi] $ and amplitude $A>0$:

\begin{equation}
\forall n \in \mathbb{Z} \text{, } x(n) =  A e^{2i \pi f_0 n + i\phi_0}.
\label{eq:sinus}
\end{equation}

The expression of the STFT is, for each frequency channel $k \in \llbracket-\frac{F-1}{2};\frac{F-1}{2}\rrbracket$ (with $F$ the odd-valued Fourier transform length) and time frame $t \in \mathbb{Z}$: 

\begin{equation}
X(k,t) = \sum_{n=0}^{N-1} x(n+tS) w(n) e^{-2i \pi \frac{k}{F} n }
\label{eq:stft}
\end{equation}
where $w$ is a $N$ sample-long analysis window and $S$ is the time shift (in samples) between successive frames. Let $W(f) = \sum_{n=0}^{N-1} w(n) e^{-2i \pi f n} $ be the discrete time Fourier transform of the analysis window for each normalized frequency $f \in [-\frac{1}{2} ; \frac{1}{2}]$. Then the STFT of the sinusoid \eqref{eq:sinus} is:

\begin{equation}
X(k,t) = A e^{2i \pi f_0 St + i\phi_0} W \left( \frac{k}{F}-f_0 \right).
\label{eq:stft_sin}
\end{equation}

%We consider an analysis window such that $w\geq0$ and $W\geq0$\footnote{Such windows can be easily constructed. For instance, if $h\geq0$ and $h_r(n)=h(-n)$, then the inequalities hold for $w=h*h_r$.}.
The unwrapped phase of the STFT is then:

\begin{equation}
\phi(k,t) = \phi_0 + 2 \pi S f_0 t + \angle W \left( \frac{k}{F}-f_0 \right)
\label{eq:phase_sinus}
\end{equation}
where $\angle z$ denotes the argument of the complex number $z$. This leads to a relationship between two successive time frames:

\begin{equation}
\phi(k,t) = \phi(k,t-1) + 2 \pi S f_0.
\label{eq:phase_sinus_update}
\end{equation}

More generally, we can compute the phase of the STFT of a frequency-modulated sinusoid. If the frequency variation is low between two successive time frames, we can generalize the previous equation:
%(i.e $\frac{df_0}{dt}(t)\ll\frac{f_0(t)}{S}$ )
%\begin{equation}
%\phi(f,t) = \phi_0 + 2 \pi S \int_0^t f_0(u) du
%\label{eq:phase_variable}
%\end{equation}
%
%We then obtain the relationship:
%
%\begin{equation}
%\phi(f,t) = \phi(f,t-1) + 2 \pi S \int_{t-1}^t f_0(u) du
%\label{eq:phase_update_variable}
%\end{equation}
%
%If $f_0(u) \approx f_0(t)$ then the previous equation becomes:

\begin{equation}
\phi(k,t) = \phi(k,t-1) + 2 \pi S f_0(t).
\label{eq:phase_update_variable_approx}
\end{equation}

Instantaneous frequency must then be estimated at each time frame to encompass variable frequency signals such as vibratos, which commonly occur in music signals (singing voice or cello signals for instance). 

\subsection{Instantaneous frequency estimation}
Quadratic interpolation FFT (QIFFT) is a powerful tool for estimating the instantaneous frequency near a magnitude peak in the spectrum~\cite{Abe2004}. It consists in approximating the shape of a spectrum near a magnitude peak by a parabola. This parabolic approximation is justified theoretically for Gaussian analysis windows, and used in practical applications for any window type. The computation of the maximum of the parabola leads to the instantaneous frequency estimate. Note that this technique is suitable for signals where only one sinusoid is active per frequency channel.

The frequency bias of this method can be reduced by increasing the zero-padding factor~\cite{Abe2004a}. For a Hann window without zero-padding, the frequency estimation error is less than $1$ \%, which is hardly perceptible in most music applications according to the authors.

\subsection{Regions of influence}
When the mixture is composed of one sinusoid, the phase must be unwrapped in all frequency channels according to \eqref{eq:phase_sinus_update} using the instantaneous frequency $f_0$. When there is more than one sinusoid, frequency estimation is performed near each magnitude peak. Then, the whole frequency range must be decomposed in several regions (\emph{regions of influence}~\cite{Laroche1999}) to ensure that the phase in a given frequency channel is unwrapped with the appropriate instantaneous frequency.

At time frame $t$, we consider a magnitude peak $A_p$ in channel $k_p$. The magnitudes (resp. the frequency channels) of neighboring peaks are denoted $A_{p-1}$ and $A_{p+1}$ (resp. $k_{p-1}$ and $k_{p+1}$). We define the region of influence $I_p$ of the $p$-th peak as follows:

\begin{equation}
I_p = \left[  \frac{A_p k_{p-1} + A_{p-1} k_p}{A_p + A_{p-1}}  ;  \frac{A_p k_{p+1} + A_{p+1} k_p}{A_p + A_{p+1}}  \right].
\label{eq:region_influence}
\end{equation}

The greater $A_p$ is relatively to $A_{p-1}$ and $A_{p+1}$, the wider $I_p$ is. Note that other definitions of regions of influence exist, such as choosing the limit between two peaks as the channel of lowest energy~\cite{Laroche1999}.

\section{Onset phase reconstruction}
\label{sec:onset}

\subsection{Impulse model}
Impulse signals are useful to obtain a relationship between phases over frequencies (vertical unwrapping)~\cite{Sugiyama2013}. Although they do not accurately model attack sounds, they provide simple equations that can be further improved for more complex signals. The model is:

\begin{equation}
\forall n \in \mathbb{Z} \text{, } x(n) = A \delta_{n-n_0},
\end{equation}
where $\delta$ is equal to one if $n=n_0$ (the so-called \emph{attack time}) and zero elsewhere and $A>0$ is the amplitude. Its STFT is equal to zero except within attack frames:

\begin{equation}
X(k,t) = A w(n_0-St) e^{-2i \pi \frac{k}{F} (n_0 - St)}.
\label{eq:stft_dirac}
\end{equation}

We can then obtain a relationship between the phases of two successive frequency channels within an onset frame, assuming that $w\geq0$:

\begin{equation}
\phi(k,t) = \phi(k-1,t) - \frac{2 \pi}{F} (n_0 - St),
\label{eq:phase_impulse_update}
\end{equation}
and $\phi(0,t)=0$. The similarity between \eqref{eq:phase_impulse_update} and \eqref{eq:phase_sinus_update} was expected because the impulse is the dual of the sinusoid in the TF domain. This comparison naturally leads to estimating parameter $n_0$ (the "instantaneous" attack time) in each frequency channel as we previously estimated $f_0$ (the instantaneous frequency) in each time frame (cf. equation \eqref{eq:phase_update_variable_approx}). This leads to the following vertical unwrapping equation:

\begin{equation}
\phi(k,t) = \phi(k-1,t) - \frac{2 \pi}{F}  (n_0(k) - St).
\label{eq:phase_impulse_update_approx}
\end{equation}

\subsection{Attack time estimation}
\label{sec:n0}

In order to estimate $n_0(k)$, we look at the magnitude of the STFT of the impulse in a frequency channel $k$:

\begin{equation}
|X(k,t)| = Aw(n_0(k) - St).
\end{equation}

We then choose $n_0$ such that the STFT magnitude of the impulse over onset frames has a shape similar to that of the analysis window. For instance, a least-squares estimation method can be used. We tested this technique on synthetic mixtures of impulses: perfect reconstruction has been reached. Alternatively, we can also estimate $n_0(k)$ with a temporal QIFFT and update the phase with \eqref{eq:phase_impulse_update_approx}.

\section{Experimental evaluation}
\label{sec:exp}

\subsection{Protocol and datasets}
\label{sec:dataset}
The MATLAB Tempogram Toolbox \cite{Grosche2011} provides a fast and reliable onset frames detection from spectrograms. We use several datasets in our experiments:

\begin{itemize}
\item[A:] 30 mixtures of piano notes from the Midi Aligned Piano Sounds (MAPS) database \cite{Emiya2010a},
\item[B:] 30 piano pieces from the MAPS database,
\item[C:] 12 string quartets from the SCore Informed Source Separation DataBase (SCISSDB) \cite{Hennequin2011a},
\item[D:] 40 speech excerpts from the Computational Hearing in Multisource Environments (CHiME) database \cite{Barker2013}.
\end{itemize}

The data is sampled at $F_s=11025$ Hz and the STFT is computed with a $512$ sample-long Hann window, $75$ \% overlap and no zero-padding. The Signal to Distortion Ratio (SDR) is used for performance measurement. It is computed with the \textbf{BSS Eval} toolbox \cite{Vincent2006} and expressed in dB. The popular consistency-based Griffin and Lim (GL) algorithm \cite{Griffin1984} is also used as a reference. We run $200$ iterations of this algorithm (performance is not further improved beyond). It is initialized with random values, except for TF bins where the phase is known. Results are averaged over $30$ initializations.

Simulations are run on a $3.60$GHz CPU processor and $16$Go RAM computer. The related MATLAB code and some sound excerpts are provided on the author web page\footnote{\url{http://perso.telecom-paristech.fr/magron/}.}.

\subsection{Horizontal phase reconstruction}
%A first experiment consists in estimating instantaneous frequencies on synthetic mixtures of damped sinusoids, which parameters (in particular the frequencies) are user-defined\footnote{We cannot fully detail this experiment here because of a lack of space, but the MATLAB code for this study is available at \url{http://perso.telecom-paristech.fr/magron/}.}. Frequency estimation error with QIFFT is below the threshold of $0.2$~\%, commonly referred to as the maximal human auditory resolution.

\begin{figure}
\centering
\includegraphics[scale=0.4]{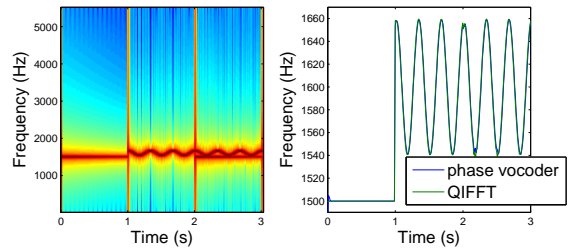}
\caption{Spectrogram of a mixture with vibrato (left) and instantaneous frequencies in the $2800$ Hz channel (right)}
\label{fig:freq_est_vibrato}
\end{figure}

Figure \ref{fig:freq_est_vibrato} illustrates the instantaneous frequencies estimated with the phase vocoder technique~\cite{Laroche1999}, used as a reference, and with our algorithm on a vibrato. Identical results are obtained. Our method is thus suitable for estimating variable instantaneous frequency signals as well as stationary components. We computed the average frequency error between phase vocoder and QIFFT estimates for the datasets presented in section \ref{sec:dataset}. The results presented in the first column of Table~\ref{tab:horiz} confirm that QIFFT provides an accurate frequency estimation.

\begin{table}
\center
\begin{tabular}{c|c||c|c}
 Dataset & Error & GL & PU \\
\hline
A & $0.38$ & $-6.9$  & $\mathbf{2.5}$ \\
\hline
B & $0.36$ & $-12.6$ & $\mathbf{1.7}$  \\
\hline
C & $0.41$ & $-9.7$ & $\mathbf{5.3}$  \\
\hline
D & $0.52$ & $-0.4$ & $\mathbf{0.5}$  \\
\hline

\end{tabular}
\caption{Frequency estimation error (\%) and reconstruction performance (SDR in dB) for various audio datasets}
\label{tab:horiz}
\end{table}

Table~\ref{tab:horiz} also presents reconstruction performance for Griffin and Lim (GL) and our Phase Unwrapping (PU) algorithms. In both cases the onset phases are known. Our approach significantly outperforms the traditional GL method: both stationary and variable frequency signals are reconstructed accurately. In addition, our algorithm is faster than the GL technique: on a $3$min $48$s piano piece, the reconstruction is performed in $18$s with our approach and in $623$s with GL algorithm.

\subsection{Onset phase reconstruction}

Onset phases can be reconstructed with $n_0$-estimation using the impulse magnitude (\textbf{Imp}) or with QIFFT (\textbf{QI}). We also test random phases values (\textbf{Rand}, no vertical coherence), zero phases (\textbf{0}, partials in phase) and alternating partial phases between $0$ and $\pi$ (\textbf{Alt}, phase-opposed partials). These choices are justified by the observation of the phase relationships between piano partials in musical acoustics~\cite{Galembo2001}. The phase of the partials is then fully recovered with horizontal unwrapping. We test these methods on dataset A. Results presented in Table \ref{tab:pitched} show that all our approaches provide better results than GL algorithm on this class of signals. Onset phase unwrapping with $n_0$-estimation based on QIFFT provides the best result, ensuring some form of vertical coherence. In particular, we perceptually observe that this approach provides a neat percussive attack.

%\begin{table}
%\centering
%\begin{tabular}{c|c}
% Method & SDR (dB) \\
% \hline
% \hline
% GL   & $-7.9$	\\
% \hline
% PU-Impulse  & $-4.0$	\\
% \hline
% PU-QIFFT  & $\mathbf{-2.6}$	\\
% \hline
% PU-Rand  & $-4.3$	\\
% \hline
% PU-0     & $-4.7$	\\
% \hline
% PU-Alt &  $-3.5$ \\
% \hline
% 
%\end{tabular}
%\caption{Signal reconstruction performance of different methods on dataset A}
%\label{tab:pitched}
%\end{table}

\begin{table}
\centering
\begin{tabular}{c||c||c|c|c|c|c}
 Method & GL & Imp & QI & Rand & 0 & Alt\\
 \hline
 SDR (dB) & $-7.9$ & $-4.0$ & $\mathbf{-2.6}$ & $-4.3$ & $-4.7$ &  $-3.5$ \\
  \hline
\end{tabular}
\caption{Signal reconstruction performance of different methods on dataset A}
\label{tab:pitched}
\end{table}

\subsection{Complete phase reconstruction}

We consider unaltered magnitude spectrograms from dataset~A. A variable percentage of the STFT phases is randomly corrupted. We evaluate the performance of our algorithm to restore the phase both on onset and non-onset frames. 

\begin{figure}
\centering
\includegraphics[scale=0.35]{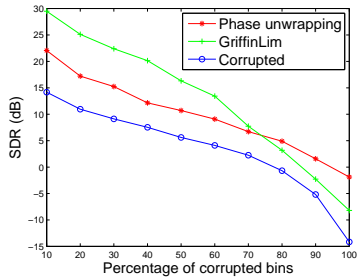}
\caption{Reconstruction performance of different methods and percentages of corruption on dataset A}
\label{fig:SDR_phase_inpainting}
\end{figure}

Figure~\ref{fig:SDR_phase_inpainting} confirms the potential of this technique. Our method produced an average increase in SDR of $6$dB over the corrupted data. It also performs better than the GL algorithm when a high percentage of the STFT phases must be recovered.

However, note that this experiment consists in phase reconstruction of \emph{consistent} spectrograms (\emph{i.e} positive matrices that are the magnitude of the STFT of a time signal): GL algorithm is then naturally advantaged in this case. Realistic applications (cf. next section) involve the restoration of both phase and magnitude, which leads to inconsistent spectrograms.

\section{Application to audio restoration}
\label{sec:inpainting}

A common alteration of music signals is the presence of noise on short time periods (a few samples) called clicks. We corrupt time signals with clicks that represent less than $1$~\% of the total duration. Clicks are obtained by differentiating a $10$ sample-long Hann window and added to the clean signal.

\begin{figure}
\centering
\includegraphics[scale=0.45]{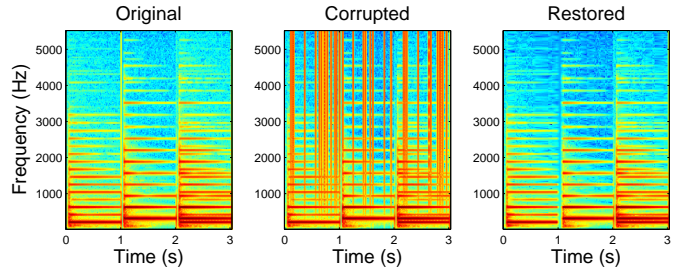}
\caption{Piano note spectrogram: original (left), corrupted (center) and restored (right)}
\label{fig:restoration_spectrogram}
\end{figure}

Magnitude restoration of missing bins is performed by linear interpolation of the log-magnitudes in each frequency channel. Figure~\ref{fig:restoration_spectrogram} illustrates this technique. Phase recovery is then performed with our method (PU) or alternatively with the GL algorithm.
We compare those results to the traditional restoration method based on autoregressive (AR) modeling of the time signal~\cite{Godsill1998}, and with HRNMF~\cite{Badeau2014}.

\begin{table}
\center
\begin{tabular}{c||c|c||c|c}

Dataset & AR & HRNMF  & GL & PU \\
\hline
A & $11.4$ & $\mathbf{16.9}$ & $8.6$ & $\mathbf{11.7}$ \\
\hline
B & $4.3$ & $\mathbf{10.9}$ & $5.9$ & $\mathbf{7.1}$  \\
\hline
C & $8.2$ & $\mathbf{10.6}$ & $6.6$ & $\mathbf{7.1}$  \\
\hline
D & $8.3$ & $\mathbf{10.9}$ & $8.9$ & $\mathbf{9.4}$  \\
\hline

\end{tabular}
\caption{Signal restoration performance (SDR in dB) for various methods and datasets}
\label{tab:restoration}
\end{table}

Table~\ref{tab:restoration} presents results of restoration. HRNMF provides the best results in terms of SDR. Though, our approach outperforms the traditional method and GL algorithm. Besides, we underline that the HRNMF model uses the phase of the non-corrupted bins, while our algorithm is blind. Lastly, our technique remains faster than HRNMF: for a $3$min$55$s piano piece, restoration is performed in $99$s with our algorithm and in $222$s with HRNMF.

\section{Conclusion}
\label{sec:conclu}

The new phase reconstruction technique introduced in this work appears to be an efficient and promising method. The analysis of mixtures of sinusoids leads to relationships between successive TF bins phases. Physical parameters such as instantaneous frequencies and attack times are estimated dynamically, encompassing a variety of signals such as piano and cellos sounds. The phase is then unwrapped in all frequency channels for onset frames and over time for partials. Experiments have demonstrated the accuracy of this method, and we integrated it in an audio restoration framework. Better results than with traditional methods have been reached.

The reconstruction of onset frames still needs to be improved as suggested by the variety of data. Further work will focus on exploiting known phase data for reconstruction: missing bins can be inferred from observed phase values. Alternatively, time-invariant parameters such as phase offsets between partials \cite{Kirchhoff2014} can be used. Such developments will be introduced in an audio source separation framework, where the phase of the mixture can be exploited.

\bibliographystyle{IEEEbib}
\bibliography{references_eusipco2015}

\begin{thebibliography}{10}

\bibitem{Smaragdis2003}
Paris Smaragdis and Judith~C. Brown,
\newblock ``Non-negative matrix factorization for polyphonic music
  transcription,''
\newblock in {\em Proc. of IEEE WASPAA}, October 2003.

\bibitem{Fitzgerald2012}
Derry Fitzgerald and Dan Barry,
\newblock ``On inpainting the adress algorithm,''
\newblock in {\em Proc. of IET ISSC}, June 2012.

\bibitem{Griffin1984}
Daniel Griffin and Jae Lim,
\newblock ``Signal estimation from modified short-time {F}ourier transform,''
\newblock {\em IEEE Transactions on Acoustics, Speech and Signal Processing},
  vol. 32, no. 2, pp. 236--243, April 1984.

\bibitem{Magron2015}
Paul Magron, Roland Badeau, and Bertrand David,
\newblock ``Phase reconstruction in {NMF} for audio source separation: An
  insightful benchmark,''
\newblock in {\em Proc. of IEEE ICASSP}, April 2015.

\bibitem{Badeau2014}
Roland Badeau and Mark~D. Plumbley,
\newblock ``Multichannel {High} {Resolution} {NMF} for modelling convolutive
  mixtures of non-stationary signals in the time-frequency domain,''
\newblock {\em IEEE Transactions on Audio Speech and Language Processing}, vol.
  22, no. 11, pp. 1670--1680, November 2014.

\bibitem{Laroche1999}
Jean Laroche and Mark Dolson,
\newblock ``Improved phase vocoder time-scale modification of audio,''
\newblock {\em IEEE Transactions on Speech and Audio Processing}, vol. 7, no.
  3, pp. 323--332, May 1999.

\bibitem{Bronson2014}
James Bronson and Philippe Depalle,
\newblock ``Phase constrained complex {NMF}: {Separating} overlapping partials
  in mixtures of harmonic musical sources,''
\newblock in {\em Proc. of IEEE ICASSP}, May 2014.

\bibitem{Krawczyk2014}
Martin Krawczyk and Timo Gerkmann,
\newblock ``{STFT} phase reconstruction in voiced speech for an improved
  single-channel speech enhancement,''
\newblock {\em IEEE/ACM Transactions on Audio, Speech, and Language
  Processing}, vol. 22, no. 12, pp. 1931--1940, December 2014.

\bibitem{Galembo2001}
Alexander Galembo, Anders Askenfelt, Lola~L. Cudy, and Franck~A. Russo,
\newblock ``Effects of relative phases on pitch and timbre in the piano bass
  range,''
\newblock {\em The Journal of the Acoustical Society of America}, vol. 110, no.
  3, pp. 1649--1666, September 2001.

\bibitem{Abe2004}
Mototsugu Abe and Julius O.~Smith III,
\newblock ``Design criteria for simple sinusoidal parameter estimation based on
  quadratic interpolation of {FFT} magnitude peaks,''
\newblock in {\em Audio Engineering Society Convention 117}. Audio Engineering
  Society, May 2004.

\bibitem{Abe2004a}
Mototsugu Abe and Julius O.~Smith III,
\newblock ``Design criteria for the quadratically interpolated {FFT} method
  (i): Bias due to interpolation,''
\newblock Tech. {R}ep. STAN-M-117, Stanford University, Department of Music,
  2004.

\bibitem{Sugiyama2013}
Akihiko Sugiyama and Ryoji Miyahara,
\newblock ``Tapping-noise suppression with magnitude-weighted phase-based
  detection,''
\newblock in {\em Proc. of IEEE WASPAA}, October 2013.

\bibitem{Grosche2011}
Peter Grosche and Meinard M{\"u}ller,
\newblock ``{T}empogram {T}oolbox: {MATLAB} tempo and pulse analysis of music
  recordings,''
\newblock in {\em Proc. of ISMIR}, October 2011.

\bibitem{Emiya2010a}
Valentin Emiya, Nancy Bertin, Bertrand David, and Roland Badeau,
\newblock ``{MAPS} - {A} piano database for multipitch estimation and automatic
  transcription of music,''
\newblock Tech. {R}ep. 2010D017, T\'el\'ecom ParisTech, Paris, France, July
  2010.

\bibitem{Hennequin2011a}
Romain Hennequin, Roland Badeau, and Bertrand David,
\newblock ``Score informed audio source separation using a parametric model of
  non-negative spectrogram,''
\newblock in {\em Proc. of IEEE ICASSP}, May 2011.

\bibitem{Barker2013}
Jon Barker, Emmanuel Vincent, Ning Ma, Heidi Christensen, and Phil Green,
\newblock ``{The PASCAL CHiME Speech Separation and Recognition Challenge},''
\newblock {\em {Computer Speech and Language}}, vol. 27, no. 3, pp. 621--633,
  Feb. 2013.

\bibitem{Vincent2006}
Emmanuel. Vincent, R\'emi Gribonval, and C\'edric F{\'e}votte,
\newblock ``Performance measurement in blind audio source separation,''
\newblock {\em IEEE Transactions on Speech and Audio Processing}, vol. 14, no.
  4, pp. 1462--1469, July 2006.

\bibitem{Godsill1998}
Simon~J. Godsill and Peter J.~W. Rayner,
\newblock {\em Digital Audio Restoration - A Statistical Model-Based Approach},
\newblock Springer-Verlag, 1998.

\bibitem{Kirchhoff2014}
Holger Kirchhoff, Roland Badeau, and Simon Dixon,
\newblock ``Towards complex matrix decomposition of spectrogram based on the
  relative phase offsets of harmonic sounds,''
\newblock in {\em Proc. of IEEE ICASSP}, May 2014.

\end{thebibliography}
\end{document}